\documentclass{article}

\usepackage{amsmath,amssymb,upgreek}
\usepackage{theorem}
\usepackage{hanmin}

\newtheorem{theorem}{Theorem}
\newtheorem{definition}{Definition}
\newtheorem{proposition}{Proposition}

\newcommand{\A}{\mathbf{A}}
\newcommand{\B}{\mathbf{B}}

\newcommand{\G}{\mathbf{H}}
\renewcommand{\H}{\mathbf{H}}
\newcommand{\K}{\mathbf{K}}

\renewcommand{\R}{\mathbf{R}}
\newcommand{\X}{\mathbf{X}}
\newcommand{\I}{\mathbf{I}}
\newcommand{\F}{\mathbf{F}}
\renewcommand{\v}{\mathbf{v}}
\newcommand{\w}{\mathbf{w}}
\newcommand{\D}{\mathrm{D}}

\author{Andrzej Hanyga\\ ul. Bitwy Warszawskiej 14/52\\
02-366 Warszawa, PL}

\title{On solutions of a class of matrix-valued convolution equations}

\begin{document}

\maketitle 

\begin{abstract}
We apply a relation between matrix-valued complete Bernstein functions and matrix-valued Stieltjes functions to prove that certain 
convolution equations 
for matrix-valued functions have unique solutions in a special class of functions. In particular the cases of the viscoelastic duality 
theorem and the Sonine equation are discussed, with applications in anisotropic linear viscoelasticity and a generalization
of fractional calculus.
\end{abstract}
\textbf{Keywords:} viscoelasticity,completely monotone, Bernstein, complete Bernstein, Stieltjes, Sonine equation, fractional calculus 

\section{Introduction}

A recent idea \cite{HanDualityNew} of simplifying the proof of a viscoelastic duality relation between the 
tensor-valued relaxation modulus and the tensor-valued creep function \cite{HanDuality}, based on a
relation between matrix-valued complete Bernstein functions and matrix-valued Stieltjes functions 
\cite{HanAnisoWaves} has led me to consider a more 
general application of this method to convolutions of matrix-valued functions, including 
the Sonine equation \cite{SamkoKilbasMarichev} for matrix-valued functions. 

Due to the similarities in the treatment 
of both equations I repeat the theorems and their proofs already published in \cite{HanDualityNew}. The last reference
is devoted to the aspects of the the duality relation relevant for viscoelasticity. In this paper I skip the details relevant 
for viscoelasticity and focus on 
proving the existence of solutions when one of the functions is either a locally integrable completely monotone (LICM) function
or a Bernstein function. 

In the case of the Sonine equation I assume that one of the functions is locally integrable and completely monotone (LICM). 
The Sonine equation was examined in much detail in \cite{SamkoCardoso}, but the authors did not assume that one of the functions
was LICM. They have constructed the inverse operator for the convolution equation $k(t)\ast x(t) = f(t)$. The inverse operator however
involves the solution $l(t)$ of the Sonine equation. Existence of such a function is the subject of our investigation 
and we prove it for $k$ in the LICM class. This problem is also studied in \cite{KochubeiArxiv}  for real-valued 
functions.

I reexamine Kochubei's suggestion that the solutions of the Sonine equation could be used to construct a 
generalization of the concepts of a derivative and integration operators along the lines of fractional calculus. In our
case it provides an "anisotropic" derivative which might turn out to be useful in the context of non-linear relaxation
equations. I also show that singularity of one of the LICM functions is essential for a correct construction of generalized 
"fractional" calculus.

\section{Convolutions equations for matrix-valued functions.}

Consider the general convolutional equation 
\begin{equation} \label{def}
\A(t)\ast\X(t) = \R(t)
\end{equation}
where $\A(t)$ and $\R(t)$ are two square matrix-valued functions defined for $t \in ]0,\infty[$ in a class to be specified, while 
$\X(t)$ is a square matrix-valued function defined be equation~\eqref{def}. We shall determine the properties of the function $\X(t)$.
The ranks of the matrices are equal and will be denoted by $m$. Let $\mathcal{S}_m$ denote the space of real square matrices of rank
$m$. 

The convolution is defined by the formula
$$\A(t)\ast \B(t) := \int_0^t \A(s)\, \B(t-s)\, \dd s.$$

\begin{definition}
A matrix-valued function $\A:\, ]0,\infty[ \rightarrow \mathcal{S}_m$ is said to be {\em completely monotone} (CM) if it is 
infinitely differentiable and for every vector
$\v \in \mathbb{R}^m$ the following inequalities are satisfied
\begin{equation} \label{CM}
\forall t > 0 \; \forall n \in \mathbb{N} \; \; (-1)^n \, \D^n \v^\mathsf{T}\, \A(t)\, \v \geq 0 
\end{equation}
\end{definition}
The above definition allows for a singularity at 0. 

\begin{definition}
A matrix-valued function $\A:\; ]0,\infty[ \rightarrow \mathcal{S}_m$ is said to be {\em locally integrable completely monotone} 
(LICM) if it is CM and integrable over $]0,1]$.
\end{definition}

\begin{definition}
A matrix-valued function $\A:\; ]0,\infty[ \rightarrow \mathcal{S}_m$ is said to be a {\em Bernstein function} if it is differentiable 
and its derivative is CM.
\end{definition}

If $\A$ is a matrix-valued Bernstein function, then for every $\v \in \mathbb{R}^m$ the function $t \rightarrow \v^\mathsf{T}\, \A(t)\, 
\v$, $\;t > 0$, is non-decreasing and continuous, hence it has a finite limit at $t = 0$. Its derivative is a LICM function.

The Laplace transform $\tilde{\A}(p)$ of a matrix-valued function $\A(t)$ is defined as usual by the formula
$$\tilde{\A}(p) := \int_0^\infty \e^{-p t} \, \A(t)\, \dd t$$
for every $p \in \mathbb{C}$ such that the integral exists. 

The Laplace transform exists for $p > 0$ and a LICM function $\A$ 
because such a function is locally integrable and non-increasing. It also exists for $p > 0$ if $\A$ is a matrix-valued Bernstein 
function because $$\int_0^\infty \e^{-p t} \int_0^t \A(s) \, \dd s = \frac{1}{p} \int_0^\infty \e^{-p t}\, \B(t)\, \dd t,$$
where $\B$ is the LICM derivative of $\A$.

\begin{proposition} \label{PCM}
If $\A$ is a symmetric matrix-valued LICM function and for each vector $\v \in \mathbb{R}^m$ the function 
$\v^\mathsf{T} \, \A(t)\, \v, t > 0$ is not identically zero, then 
the matrix $\tilde{\A}(p)$ is invertible for every $p > 0$.
\end{proposition}
\noindent\textbf{Proof.}

For each non-zero $\v \in \mathbb{R}^m$ there is a real $t_1(\v) > 0$ such that $\v^\mathsf{T}\, \A(t_1(\v)) \, \v > 0$. The function
$\v^\mathsf{T}\, \A(t)\, \v$ is continuous, hence it is positive on some interval $\mathcal{I} \subset ]0,\infty[$, while it is non-negative on $]0,\infty[$. Hence for every non-zero $\v \in \mathbb{R}^m$, $p > 0$ we have $\v^\mathsf{T} \,\tilde{\A}(p)\, \v > 0$. 
The matrix $\A(p)$ is symmetric and positive definite, hence
it is invertible.
\mbox{ }\hfil$\Box$

\begin{proposition} \label{PBF}
If $\A$ is a symmetric matrix-valued Bernstein function and for each vector $\v \in \mathbb{R}^m$ the function 
$\v^\mathsf{T} \, \A(t)\, \v, t > 0$ is not identically zero, then 
the matrix $\tilde{\A}(p)$ is invertible for every $p > 0$.
\end{proposition}
The proof is analogous to the previous proposition.

It is easy to check the familiar identity
$$\int_0^\infty \e^{-p t}\, \A(t)\ast\B(t) \, \dd t = \tilde{\A}(p)\, \tilde{\B}(p), \; p > 0$$
provided both Laplace transforms on the right-hand side exist.
Consequently equation~\eqref{def} implies the equation
\begin{equation} \label{Lap}
\tilde{\A}(p) \, \tilde{\X}(p) = \tilde{\R}(p)
\end{equation}

If the matrix $\tilde{\A}(p)$ is invertible for $p > 0$, then the unique solution of \eqref{Lap} is 
\begin{equation} \label{Lap1}
\tilde{\X}(p) = \tilde{\A}(p)^{-1}\, \tilde{\R}(p), \;\; p > 0
\end{equation}

This is in particular true if $\A$ is an $\mathcal{S}_+$-valued LICM function. For $\R(t) = t \, \I, t > 0$,
where $\I$ is the identity operator on $\mathbb{R}^m$,  we have $\tilde{\R}(p) = p^{-2}\, \I$ and
\begin{equation}
p \, \tilde{\X}(p) = \left[ p\, \tilde{\A}(p) \right]^{-1}
\end{equation}
We shall show that under these assumptions the solution $\X$ of equation~\eqref{def} is an $\mathcal{S}_+$-valued Bernstein function.

Similarly, if $\A$ is an $\mathcal{S}_+$-valued BF and $\R(t) = t \, \I, t > 0$, then $\X$ is an $\mathcal{S}_+$-valued LICM function.

By transposition the results obtained below also apply to equations of the form
$$\X(t)\ast \A(t) = \R(t)$$.

\section{Main theorems and their proofs.}

\begin{theorem} \label{thm1}
If $\A$ is an $\mathcal{S}_+$-valued LICM function and \\ \vspace{0.2cm}
\parbox{\textwidth}{$(\ast)$ for each non-zero vector $\v \in \mathbb{R}^m$ the function 
$\v^\mathsf{T} \, \A(t)\, \v, t > 0$, is not identically zero,} \\ 
then equation~\eqref{def} with $\R(t) = t\, \I$,
has a unique solution $\X$, which is a $\mathcal{S}_+$-valued Bernstein function.
\end{theorem}

\begin{theorem} \label{thm2}
If $\A$ is an $\mathcal{S}_+$-valued Bernstein function satisfying Condition~$(\ast)$, 
 then equation~\eqref{def} with 
 $\R(t) = t\, \I$, has a unique solution $\X$, 
\begin{equation} \label{uu}
\X(t) = \B\, \delta(t) + \mathbf{F}(t)
\end{equation}
where $\B \in \mathcal{S}_+$ and $\mathbf{F}$ is a $\mathcal{S}_+$-valued LICM function.
\end{theorem}

It follows from Theorems~\ref{thm1} and \ref{thm2} that for every LICM function $\kappa(t)$ the equation 
$\kappa\ast \lambda = t$ for $t \geq 0$ has a unique solution $\lambda$. The solution function $\lambda$  is a BF. 
For every
 $\kappa$ which is a BF the same equation has a solution $\lambda$ which is a LICM function plus a Dirac delta term. 

If $\kappa$ and $\lambda$ 
satisfy the same equation but $\kappa$ is neither a BF nor LICM then the same is true for $\lambda$. Do such 
pairs exist? Here is an example of such a pair of which none is CM nor BF:
$\kappa(t) = t^{1/2}\, I_\nu(2 t^{1/2})$ and $\lambda(t) = t^{-1/2}\, J_{-\nu}(2 t^{1/2})$ for $\nu > -1$.
Indeed, according to (18) 
on p.~197 and (30) on p.~185 of \cite{BatemanProject} 
$\tilde{\kappa}(p) = p^{-\nu-1} \, \exp(1/p)$ and $\tilde{\lambda}(p) = p^{\nu - 1}\, \exp{-a/p}$. 
$\lambda$ changes sign hence it is neither LICM nor a BF.\\ \vspace{0.2cm}

\noindent\textbf{Proof of Theorem~\ref{thm1}.}

On account of  \eqref{mlapLICM}
$$p \,\tilde{\mathbf{A}}(p) = p \, \mathbf{C} + p \int_{[0,\infty[} (p + r)^{-1}\, \mathbf{G}(r) \, \mu(\dd r),$$
where $\mathbf{C} \geq 0$,
hence $p \, \tilde{\A}(p)$ is an  $\mathcal{S}_+$-valued CBF. 

On account of $(\ast)$ and Proposition~\ref{PCM} the matrix $\tilde{\A}(p)$ has an inverse for $p > 0$. Hence the function 
$p \, \tilde{\A}(p)$ does not vanish identically and according to Theorem~\ref{inv} 
the inverse $p\, \tilde{\X}(p)$ of $p\, \tilde{\A}(p)$ is an $\mathcal{S}_+$-valued Stieltjes function and has the form
\begin{equation} \label{yu}
\mathbf{B} + p^{-1}\, \mathbf{D} + \int_{]0,\infty[} (p + r)^{-1}\,\mathbf{H}(r)\, \mu(\dd r),
\end{equation}
where $\mathbf{D}, \mathbf{B} \in \mathcal{S}_+$, $\mu$ is a Borel measure on $]0,\infty[$ satisfying \eqref{x2} and 
$\H$ is a bounded measurable $\mathcal{S}_+$-valued function 
defined $\mu$-almost everywhere on $]0,\infty[$.
On account of equation~\eqref{Lap1} $p \, \tilde{\mathbf{X}}(p)$ has the form given by equation~\eqref{yu}. It follows that  
$$\mathbf{X}(t) = \mathbf{B} + t\, \mathbf{D} + \int_0^t \mathbf{K}(s) \, \dd s, $$
where $$\mathbf{K}(t) := \int_{]0,\infty[} \e^{-r t}\, \H(r)\, \mu(\dd r)$$ is a LICM. It follows that $\mathbf{X}$ is
an $\mathcal{S}_+$-valued Bernstein function.

\mbox{ }\hfil$\Box$

\noindent\textbf{Proof of Theorem~\ref{thm2}.}

Condition $(\ast)$ and Proposition~\ref{PBF} ensures that the matrix $\tilde{\A}(p)$  is invertible for $p \geq 0$.

We now note that $\mathbf{A}(t) = \int_0^t \mathbf{L}(s)\, \dd s$, where $\mathbf{L}$ is an $\mathcal{S}_+$-valued LICM function.
 By the Bochner Theorem there is a Borel measure $\mu$ satisfying equation~\eqref{x2} and an $\mathcal{S}_+$-valued function
$\mathbf{G}(r)$, $r \geq 0$, bounded everywhere except perhaps on a set of $\mu$ measure zero, such that 
$$\mathbf{L}(t) = \int_{[0,\infty]} \e^{-r s} \, \mathbf{G}(r) \, \mu(\dd r).$$
Consequently $p \, \tilde{\A}(p) = \tilde{\mathbf{L}}(p) = \int_{[0,\infty[} (p + s)^{-1}\, \mathbf{G}(r) 
\mu(\dd r)$ is an $\mathcal{S}_+$-valued Stiel\-tjes function and it is not identically vanishing.
By Theorem~\ref{inv} its inverse is an
$\mathcal{S}_+$-valued CBF and therefore it has the form
$$p\, \mathbf{B} + p \int_{[0,\infty[} (p + r)^{-1} \, \mathbf{G}(r)\, \mu(\dd r),$$
for some  $\B \in \mathcal{S}_+$, a Borel measure $\mu$ satisfying \eqref{x2} and a 
measurable $\mathcal{S}_+$-valued function $\mathbf{G}$ bounded $\mu$-almost everywhere on $]0,\infty[$.

Equation~\eqref{def} is satisfied if 
$\X$ is given by equation~\eqref{uu} with $$\mathbf{F}(t) :=  
\int_{[0,\infty[} \e^{-r t} \, \mathbf{G}(r)\, \mu(\dd r).$$\\ 

\mbox{ }\hfil $\Box$

\begin{theorem} \label{thm3}
If $\A$ is a non-zero $\mathcal{S}_+$-valued LICM function and 
the limit $\A_0 := \lim_{t\rightarrow 0} \A(t)^{-1}$ exists, then equation~\eqref{def} with $\R(t) = \I$
has a unique solution $\X = \A_0 \, \delta(t) + \F(t)$,  where $\F$ is an $\mathcal{S}_+$-valued LICM function. 
\end{theorem}

The equation $k\ast l = 1$ in for locally integrable real-valued functions $k$ and $l$ is known as the Sonine equation 
\cite{SamkoKilbasMarichev}. If for a given $k \in \mathcal{L}^1_{\mathrm{loc}}([0,\infty[)$ there is an 
$l \in \mathcal{L}^1_{\mathrm{loc}}([0,\infty[)$ satisfying the above equation, then $k$ is called a {\em Sonine kernel}, 
while $k, l$ are known as a {\em Sonine pair}. Sonine pairs are studied in some detail in \cite{SamkoCardoso}. 
Theorem~\ref{thm3} asserts in particular that every LICM function or matrix-valued LICM function is a Sonine kernel and 
in this case the Sonine pair consists of two LICM functions.
For real-valued functions this fact has apparently been discovered by Kochubei \cite{KochubeiArxiv}.

However, not every Sonine pair consists of LICM functions. An counterexample is the Sonine pair $k_\lambda(t) := t^{-\lambda/2}\,
J_{-\lambda}(2 t^{1/2})$ with the Laplace transform $\tilde{k}_\lambda(p) = \exp(-1/p) \, p^{\lambda-1}$ 
(\cite{BatemanProject} p. 185 (30)) and
$l_\lambda(t) t^{(\lambda-1)/2} \, I_{\lambda-1}(2 t^{1/2})$ with $\tilde{l}_\lambda(p) = \exp(1/p)\, p^{-\lambda-2}$
(\cite{BatemanProject} p. 197 (18))
for $\lambda > 0$. The function $k_\lambda$ changes sign and therefore is not CM, for example
$k_{1/2}(t) = \sqrt{2/\uppi} \, \cos(2 t^{1/2})/t^{3/4}$. 

The following matrix-valued Sonine pairs are of particular interest:
\begin{enumerate}
\item $k(t) \, \K_0$ and $l(t)\, \K_0^{\;-1}$, where $k, l$ are a Sonine pair of LICM functions;
\item $\mathrm{diag} \{k_n(t), n=1,... m\}$ and $\mathrm{diag} \{l_n(t), n=1,... m\}$, where 
$(k_n, l_n)$ are Sonine pairs of LICM functions for $n = 1, \ldots, m$.
\end{enumerate}

Many CM functions are known \cite{MillerSamko01}, but it is often more difficult to find the other member of the Sonine pair.
The simplest Sonine pair of CM functions is $k(t) = t^{\alpha-1}/\Gamma(\alpha)$ and $l(t) = t^{-\alpha}/\Gamma(1-\alpha)$,
$0 < \alpha < 1$. Using the Laplace transforms $\tilde{k}(p) = (p + \lambda)^{-\alpha}$ and  
$\mathcal{L}[\Gamma(-\alpha,\lambda t)](p) = \Gamma(-\alpha)\, \lambda^{-\alpha} \left[\lambda^\alpha - (\lambda + p)^\alpha\right]/p$
one gets another pair  $k(t) = t^{\alpha-1}\, \e^{-\lambda t}/\Gamma(\alpha)$, 
$\lambda > 0$, with $l(t) = \lambda^\alpha\, \times \\ \left[1 - \Gamma(-\alpha,\lambda t)/\Gamma(-\alpha)\right]$,
$\lambda \geq 0$, $0 < \alpha < 1$. 

It is also interesting that for an arbitrary analytic function $k(t)$ there is another analytic function $l(t)$ such that
$k(t)\, t^{\alpha-1}/\Gamma(\alpha)$ and $l(t) \, t^{-\alpha}/\Gamma(1-\alpha)$ are a Sonine pair and there is an algorithm for 
calculating the power series of $l(t)$ given the power series for $k(t)$ \cite{SamkoCardoso, Wick}.\\
\vspace{0.2cm}

\noindent\textbf{Proof of Theorem~\ref{thm3}.}

The Laplace transform $\tilde{\A}(p)$ is a symmetric positive definite matrix for every  $p > 0$.
Equation~\eqref{def} is equivalent to $\tilde{X}(p) = \left[p\, \tilde{\A}(p)\right]^{-1}$ 
The right-hand side is the algebraic inverse of a matrix-valued CBF, hence it is a Stieltjes function of 
the form 
\begin{equation}  \label{eq5}
 \mathbf{H}(p) = \mathbf{C} + \int_{[0,\infty[} (p + r)^{-1}\, \mathbf{G}(r)\, \mu(\dd r).
\end{equation}
where $\mathbf{C} \geq 0$, $\mathbf{G}$ is a measurable symmetric matrix-valued function bounded $\mu$-almost everywhere 
and $\mu$ is a Borel measure satisfying inequality~\eqref{x2}.

In view on inequality~\eqref{x2} the Lebesgue Dominated Convergence Theorem implies that 
$$\mathbf{C} = \lim_{p \rightarrow \infty} \mathbf{H}(p) = \left[ \lim_{p \rightarrow \infty}\,(p\, \tilde{\A}(p))\right]^{-1} 
= \A_0.$$ 

The second term on the right-hand side of equation~\eqref{eq5} is a Laplace transform 
$\tilde{\F}(p)$ of the LICM function 
\begin{equation} \label{FLICM}
\F(t) := \int_{[0,\infty[} \e^{-r t}\, \mathbf{G}(r) \, \mu(\dd r)
\end{equation}
Inverting the Laplace transformation we conclude that $\X(t) = \A_0 \, \delta(t) + \F(t)$.
\mbox{ }\hfil$\Box$

The first application of this result is the solution of a convolution equation
\begin{equation}
\A(t)\ast \v(t) = \mathbf{f}(t)
\end{equation}
Since $\X(t)\ast \A(t) = \I$, 
$$\int_0^t \v(t)\, \dd t  = \X(t)\ast \mathbf{f}(t)$$
and therefore
\begin{equation}
\v(t) = \D [\X\ast \mathbf{f}(t)] 
\end{equation}

The LICM function $\A(t)$ can have a singularity at 0 such that for every $\v \in \mathbb{R}^m$ the limit 
$\lim_{t \rightarrow 0} \, \v^\mathsf{T}\,\A(t)\, \v = \infty$. It then follows that  $\A_0 = 0$. In this case we define
the generalized Caputo $\A$-derivative by the formula
\begin{equation}
\D_\A \, \v(t) = \D \left[ \A(t)\ast \v(t) \right] - \A(t)\, \v(0) \equiv \A \ast \v^\prime 
\end{equation}
for every  absolutely continuous function $\v:\; [0,\infty[ \rightarrow \mathbb{R}^m$.
The term "derivative" is justified if the function $\A$ is singular at 0. If $\X = \F$ is the solution 
of the convolution equation~\eqref{def} with $\R(t) = \I$, then the  $\A$-integral operator is defined by the formula
\begin{equation}
\mathrm{J}_\A \, \v(t) := \F\ast \mathbf{1}\ast \v(t)
\end{equation}
where $\mathbf{1}\ast \v(t) = \int_0^t \v(s)\, \dd s$.

We then have 
\begin{theorem} \label{thm4}
Let $\A_0 = 0$.\\
The following relations hold 
\begin{eqnarray}
\mathrm{J}_\A\, \D_\A \, \w(t) = \w(t) - \w(0) \text{   for } \w \in AC([0,\infty[)\\
\D_\A\, \mathrm{J}_\A \, \v(t) = \v(t) \text{   for } \v \in \mathcal{L}^1_\mathrm{loc}([0,\infty[)
\end{eqnarray}
\end{theorem}
\noindent\textbf{Proof}\\
(1) The identity $\F \ast \A = \mathbf{1}$ implies that 
$$(\mathrm{J}_\A \, \D_\A \, \w)(t) = \int_0^t \F(s) \left. \frac{\dd}{\dd \tau} \int_0^\tau \A(\tau - r) \w(r)\, \dd r 
\right|_{\tau=t-s} \, \dd s - \w(0)$$
The first term equals
$$\int_0^t \F(s) \frac{\dd}{\dd t} \int_0^{t-s} \A(t -s - r) \, \w(r)\, \dd r\, \dd s = \frac{\dd}{\dd t} (\F\ast \A \ast \w)(t) =
\frac{\dd}{\dd t} (\mathbf{1}\ast \w)(t) = \w(t)$$
q.e.d.

\noindent (2) Let $\w := \F \ast \v$.
On account of the identity $\A \ast \F = \mathbf{1}$ 
$$\D_\A\, \mathrm{J}_\A\, \v = \frac{\dd}{\dd t} (\A\ast \F\ast \v)(t) - \A(t)\, \w(0) = \v(t) - \A(t)\, \w(0)$$

It remains to prove that $\w(0) = 0$. $\F$ is a LICM function, hence it has the form \eqref{FLICM} with $\mu$ satisfying 
equation~\eqref{x2} and 
$\vert \G(r)\, \vert \leq 1$. Hence 
$$\vert \w(t) \vert \leq \left[ \int_0^t \int_{[0,\infty[} \e^{-r s} \, \mu(\dd r)\, \dd s\right] \,\int_0^t |\v(t - s)| \, \dd s$$

For $t \leq 1$ the second factor is bounded from above by a constant 
$$\int_0^1 | \v(s)| \, \dd s  < \infty$$
because $\v$ is assumed locally integrable.
The first factor equals
\begin{equation} \label{f1}
\int_{[0,\infty[} \frac{1 - \e^{-r t}}{r} \, \mu(\dd r)
\end{equation}
From the inequality $\e^x - 1 \leq x\, \e^x$ ($x \geq 0$) follows  the inequality $1 - \e^{-x} \leq x$. We shall apply this inequality for $r \in [0,\infty[$, noting that 
$\mu([0,1]) < \infty$ because of \eqref{x2} with the inequality $1 \leq 2/(1 + r)$ valid for $r \leq 1$. For $r > 1$ we shall note that 
$1/r \leq 2/(1 + r)$. Hence expression~\eqref{f1} is bounded by 
$$t  \, \mu([0,1]) + 2 \int_{]1,\infty[} \left( 1 - \e^{-r t}\right)\, (1 + r)^{-1} \, \mu(\dd r),$$
which tends to 0 on account of \eqref{x2} and the Lebesgue Dominated Convergence Theorem. Thus $\w(0) = 0$ and the theorem has been proved.

\mbox{ }\hfil$\Box$\\

The new derivative concept provides a new approach to modeling stress relaxation in anisotropic and non-linear viscoelastic media.
A possible relaxation equation could have the form 
\begin{equation}
\D_\A \, \sigma = \mathbf{K}(\sigma,\epsilon)
\end{equation}

Theorem~\ref{thm4} applies only to (weakly) singular kernels $\A$. In particular it does not apply to Caputo-Fabrizio \cite{CF} and 
Atangana-Baleanu \cite{AB} fractional derivatives. In these cases $\A$ is non-singular but the corresponding $\X$ is no longer a superposition of Newtonian viscosity and a LICM relaxation function. There is no integral associated with these derivatives. For the Caputo-Fabrizio 
derivative we have $\A(t) = \exp(-\alpha \, t/(1- \alpha))/(1 - \alpha)$ and
$\F = (1 - \alpha)\, \delta^\prime + \delta$. In the Atangana-Baleanu case $\F$ involves $\delta^\prime$ and a strongly 
singular kernel.

\section{Conclusions.}

We have demonstrated a particular role of LICM and Bernstein kernels in two classes of convolution equations and the utility
of the concepts of CBFs and Stieltjes derivatives in the study of existence problems for these equations.

\bibliography{ownnew12,mathnew12}

\begin{thebibliography}{1}

\bibitem{AB}
A. Atangana and D. Baleanu.
\newblock New fractional derivatives with nonlocal and non-singular kernel: Theory and applications
to heat transfer model. 
\newblock {\em arxiv:} 1604.0340, 2016.

\bibitem{BatemanProject}
  A. Erd\'{e}lyi, W. Magnus, F. Oberhettinger and F. Tricomi.
  \newblock Higher {T}ranscendental {F}unctions.
  \newblock McGraw-Hill, New York, 1953

\bibitem{CF}
 M.Caputo and M.Fabrizio, 
\newblock A new definition of fractional derivative without singular kernel.
\newblock {\em Progr. Fract. Differ. Appl.},
1:73–85, 2015.

\bibitem{HanDuality}
A.~Hanyga and M.~Seredy\'{n}ska.
\newblock Relations between relaxation modulus and creep compliance in
  anisotropic linear viscoelasticity.
\newblock {\em J. of Elasticity}, 88:41--61, 2007.

\bibitem{HanAnisoWaves}
A. Hanyga 2016  Wave propagation in anisotropic elasticity. {\em J. of
  Elasticity} \textbf{122}, 231--254.

\bibitem{HanDualityNew}
A. Hanyga  2018 A simple proof of a duality theorem with applications in scalar and anisotropic viscoelasticity,
 {\em arxiv:}, 1805.07275, 2018.

\bibitem{KochubeiArxiv}
A.~N. Kochubei.
\newblock General fractional calculus, evolution equations and renewal
  processes.
\newblock {\em arxiv:}, 1105.1239, 2011.

\bibitem{MillerSamko01}
K.~S. Miller and S.~G. Samko.
\newblock Completely monotonic functions.
\newblock {\em Integr. Transf. and Spec. Fun.}, 12:389--402, 2001.

\bibitem{SamkoCardoso}
S.~G. Samko and R.~P. Cardoso.
\newblock Integral equations of the first kind of Sonine type.
\newblock {\em Intern. J. Math. and Math. Sci.}, 57:3609--3632, 2003.

\bibitem{SamkoKilbasMarichev}
S.~G. Samko, A.~A. Kilbas, and O.~I. Marichev.
\newblock {\em Fractional integrals and derivatives. Theory and applications}.
\newblock Gordon and Breach, London, 1993.

\bibitem{Wick}
J. Wick.
\newblock \"{U}ber eine Integralgleichung vom Abelschen Typ.
\newblock {\em ZAMM} \textbf{48}, 39--41, 1968.


\end{thebibliography}

\appendix
\section{Matrix-valued Stieltjes functions and Complete Bernstein functions.}\label{appAniso}

Let $\mathcal{S}_+$ denote the set of non-negative symmetric matrices.

An $\mathcal{S}_+$-valued 
function $\mathbf{A}(t)$ is CM if 
$$ (-1)^n \, \D^n \,\mathbf{A}(t) \geq 0 \hspace{1cm} \text{for} \; n = 0, 1, 2\ldots $$
where $\mathbf{B} \geq 0$ is equivalent to $\mathbf{v}^\mathsf{T}\, \mathbf{M}\, \mathbf{v} \geq 0$ for every 
$\mathbf{v} \in \mathbb{R}^6$.

The function $\mathbf{A}(t)$ is LICM if it is CM and locally integrable. 

For every $\mathcal{S}_+$-valued LICM function $\mathbf{A}$ there is a Borel measure $\mu$ 
satisfying the inequality
\begin{equation} \label{x2}
\int_{]0,\infty[} (1 + s)^{-1} \,\mu(\dd s) < \infty
\end{equation}
and an $\mathcal{S}_+$-valued 
function $\mathbf{G}$ on $[0,\infty[$ bounded on everywhere except for a set of $mu$ measure zero  such that
\begin{equation} \label{mLICM}
\mathbf{A}(t) = \int_{[0,\infty[} \e^{-r t}\, \mathbf{G}(r) \, \mu(\dd r)
\end{equation}

If $\mu(\{0\}) > 0$, then $\mathbf{G}(0)$ is defined and \eqref{mLICM} can be recast in the form
\begin{equation}\label{mLICM1}
\mathbf{A}(t) = \mathbf{B} + \int_{]0,\infty[} \e^{-r t}\, \mathbf{G}(r) \, \mu(\dd r)
\end{equation}
where $\mathbf{B} := \mu(\{0\})\, \mathbf{G}(0)$ is a positive semi-definite symmetric matrix. If $\mu(\{0\}) = 0$ then $\mathbf{B} = 0$.

The Laplace transform of the $\mathcal{S}_+$-valued LICM $\mathbf{A}(t)$ is given by the equation 
\begin{equation} \label{mlapLICM}
\tilde{\mathbf{A}}(p) = \int_{[0,\infty[} (p + r)^{-1} \, \mathbf{G}(r) \, \mu(\dd r)
\end{equation}

An $\mathcal{S}_+$-valued Bernstein function is an indefinite integral of a $\mathcal{S}_+$-valued LICM function.

We shall now recall some results from Appendix~B of \cite{HanAnisoWaves}.

A matrix-valued Stieltjes function $\mathbf{Y}(p)$ has the following integral representation:
\begin{eqnarray}\label{mStieltjes}
\mathbf{Y}(p) = \mathbf{B} + \int_{[0,\infty[} (p + r)^{-1} \, \mathbf{G}(r) \, \mu(\dd r) = \\
= \mathbf{B} + p^{-1} \mathbf{D} + \int_{]0,\infty[} (p + r)^{-1} \, \mathbf{G}(r) \, \mu(\dd r)
\end{eqnarray}
where $\mathbf{B} \in \mathcal{S}_+$, $\mu$ is a Borel measure on $]0,\infty[$ satisfying \eqref{x2} and 
 $\mathbf{G}(r)$ is an $\mathcal{S}_+$-valued function defined
$\mu$-almost everywhere on $]0,\infty[$ and $\mathbf{D} = \mu(\{0\}) \,\mathbf{G}(0)$. \\
Conversely, any matrix-valued function with the integral representation 
\eqref{mStieltjes} is an $\mathcal{S}_+$-valued Stieltjes function.

An $\mathcal{S}_+$-valued CBF $\mathbf{Z}(p)$ has the following integral representation:
\begin{eqnarray}\label{mCBF}
\mathbf{Z}(p) = p\, \mathbf{B} + p \int_{[0,\infty[} (p + r)^{-1} \, \mathbf{H}(r) \, \nu(\dd r) = \\
= \mathbf{D} + p\, \mathbf{B} + p \int_{]0,\infty[} (p + r)^{-1} \, \mathbf{H}(r) \, \nu(\dd r)
\end{eqnarray}
where $\mathbf{B}, \mathbf{D} \in \mathcal{S}_+$, $\nu$ is a Borel measure on $]0,\infty[$ satisfying \eqref{x2} and 
 $\mathbf{H}(r)$ is an $\mathcal{S}_+$-valued function defined
$\nu$-almost everywhere on $]0,\infty[$, $\mathbf{D} = \mathbf{H}(0)\, \mu(\{0\})$.\\
Conversely, any matrix-valued function with the integral representation 
\eqref{mCBF} is a $\mathcal{S}_+$-valued CBF.

It follows immediately that the the function $p^{-1} \, \mathbf{Z}(p)$, where $\mathbf{Z}$ is an $\mathcal{S}_+$-valued 
CBF function, is an $\mathcal{S}_+$-valued Stieltjes function. 

We quote Lemma~3 op. cit. in the form of the following theorem 
\begin{theorem} \label{inv}
If $\mathbf{Z}(p)$ is an $\mathcal{S}_+$-valued CBF and does not vanish identically, then $\mathbf{Z}(p)^{-1}$ is an 
$\mathcal{S}_+$-valued Stieltjes function.

Conversely, if $\mathbf{Y}(p)$ is an $\mathcal{S}_+$-valued function does not
vanish identically then $\mathbf{Y}(p)^{-1}$ is a CBF.
\end{theorem}

\end{document}